\documentclass[review]{elsarticle}

\usepackage{lineno,hyperref}
\modulolinenumbers[5]

\journal{Physics Letters A}

\usepackage{amsmath}
\usepackage{dsfont}
\begin{document}

\begin{frontmatter}

\title{Rescaling the nonadditivity parameter in Tsallis thermostatistics}

%% Group authors per affiliation:
\author[1,2]{Jan Korbel}
\address[1]{Department of Physics, Zhejiang University, Hangzhou 310027, P.~R.~China}
\address[2]{Faculty of Nuclear Sciences and Physical Engineering, Czech Technical University in Prague, B\v{r}ehov\'{a} 7, 115 19, Prague, Czech Republic}
\ead{korbeja2@fjfi.cvut.cz}

\begin{abstract}
The paper introduces nonadditivity parameter transformation group induced by Tsallis entropy. We discuss simple physical applications of a system in the contact with finite heat bath or with temperature fluctuations. With help of the transformation, it is possible to introduce generalized distributive rule in $q$-deformed algebra. We focus on MaxEnt distributions of Tsallis entropy with rescaled nonadditivity parameter under escort energy constraints. We show that each group element corresponds to one class of $q$-deformed distributions. Finally, we briefly discuss the application of the transformation to Jizba-Arimitsu hybrid entropy and its connection to Average Hybrid entropy.
\end{abstract}


\end{frontmatter}

\nolinenumbers

\section{Introduction}
Nonadditive thermodynamics is a concept generalizing ordinary additive Boltzman-Gibbs statistical physics based on Shannon entropy \cite{shannon48}. The first additive generalization of Shannon entropy was discovered by R\'{e}nyi \cite{renyi76}, who showed that the solution of additive Khichin axioms is the one-parametric class of entropies, including Shannon entropy as a special case. The first nonadditive entropy was described by Tsallis \cite{tsallis88} and its axiomatic definition was discussed e.g. in \cite{abe97}. Interestingly, the same entropy functional had been discussed before by Havrda and Charv\'{a}t in connection with information theory \cite{havrda67}. Since that, there have been introduced many other generalizations of Shanonn entropy \cite{frank20,sharma78,bercher12,arimitsu04}. Additionally, there were done several successful classifications of generalized entropies taking into account different aspects of generalized statistics \cite{hanel11,tempesta11,ilic14,biro15,jizba16}. On the other hand, Tsallis entropy represents the most popular example of nonadditive entropy with many applications in statistical physics \cite{plastino94,adib03,kaniadakis01,plastino14,jizba17}, abstract algebra \cite{borges04,nivanen03,kalogeropoulos05}, information theory \cite{bercher2012a,bercher13} or statistics \cite{cankaya17}.

In recent years, several attempts on mixing of additivity and nonadditivity effects were studied. To these effects belong crossover between ordinary gaussian distributions and $q$-gaussian distributions \cite{sicuro2016,bagchi17}. We focus on situations, when a system is described by Tsallis entropy, but the strength of nonadditivity can change. This situation can be represented by a simple example of a system in contact with finite heat bath \cite{plastino94,adib03} with rescaled number of particles in the bath or a system with temperature fluctuations \cite{beck02,biro14}. More generally, this corresponds to the situation when we have a system with polynomially growing state space \cite{hanel11,thurner11} and shift its characteristic scaling exponent. Rescaling of nonadditivity parameter brings about some non-trivial consequences which are also discussed in this paper. The rest of the paper is organized as follows: section \ref{sec:ts} defines the transformation group of nonadditivity parameter and discusses its main properties. Section \ref{sec:ap} discusses simple physical applications of the transform. Section \ref{sec:qa} describes applications in $q$-deformed algebra and presents generalized distributive laws. In section \ref{sec:me} are calculated MaxEnt distributions (obtained from Maximum entropy procedure) corresponding to Tsallis entropy with rescaled nonadditivity parameter under escort energy constraints. Section \ref{sec:ja} discusses the connection of the transformation to Jizba-Arimitsu hybrid entropy, which follows $q$-additivity rule for independent events, similarly to Tsallis entropy. The last section is devoted to conclusions.

\section{Rescaling of Tsallis nonadditivity parameter}\label{sec:ts}
Nonadditive statistical physics has been first described by Tsallis \cite{tsallis88}. He introduced the generalized entropy functional
\begin{equation}
S_q(P) = \frac{\sum_i p_i^q -1}{1-q}
\end{equation}
where $q$ plays the role of nonadditivity parameter. For $q=1$ the entropy becomes additive Shannon entropy. The nonadditivity of Tsallis entropy is for independent events $A$, $B$ expressed by the axiom
\begin{equation}\label{eq:tsallis}
S_q(P_{A \cup B}) = S_q(P_A) + S_q(P_B) + (1-q) S_q(P_A)S_q(P_B)\,
\end{equation}
where $P_A$ is the probability distribution corresponding to $A$, and similarly with $B$. One of the possible sources of nonadditivity can arise from the fact that a system is in contact with a finite heat bath. In this case, it is possible show that for a bath consisting of $N$ particles the nonadditivity parameter is defined as $q(N) = \frac{N}{N-1}$ \cite{plastino94,adib03}, and the state space grows polynomially. The polynomial growth of states is typical for systems described by Tsallis entropy, because the entropy is extensive for these systems \cite{hanel11}. This is typical for systems where a fraction of states is frozen \cite{thurner11}. The important question is what happens with the system when we change the strength of non-additive interaction. This means that we rescale the nonadditivity term in Eq. \ref{eq:tsallis}, i.e. we replace $(1-q)$ by $(q-1)/\alpha$ for some $\alpha >0$. In this case, we get a new nonadditivity parameter $q_\alpha$, which is defined by the relation
\begin{equation}
(q_\alpha - 1) = \frac{q-1}{\alpha} \qquad \Rightarrow \qquad q_\alpha = \frac{q+\alpha-1}{\alpha}\, .
\end{equation}
Interestingly, this class of nonadditivity parameter transformations conforms a one-dimensional group. It is straightforward to show the main properties of the group, which are
\begin{itemize}
  \item composition rule: $(q_{\alpha})_\beta = (q_\beta)_\alpha = q_{\alpha \cdot \beta}$
  \item associativity: $(q_{\alpha})_{\beta \cdot \gamma} = (q_{\alpha \cdot \beta})_{  \gamma} = q_{\alpha \cdot \beta \cdot \gamma}$
  \item neutral element: $q_1 = q$
  \item transformation invariant: $1_\alpha \equiv 1$.
\end{itemize}

Naturally, it is possible to think about extension of the transform beyond the region $\alpha > 0$. Actually, the region $\alpha > 0$ rescales the distance of $q$ from $1$, but does not change the sign of $(1-q)$. It means that the transform keeps $q_\alpha > 1$ for $q > 1$ and vice versa. Because of multiplicative properties of the transform, the most important is the extension to $\alpha=-1$. With this, we get
\begin{equation}
q_{-1}= \frac{q-1-1}{-1} = 2-q
\end{equation}
which is the well-known additive duality of Tsallis entropy. Unfortunately, this transformation can lead to negative values of $q_{-1}$, which is usually unwanted because Tsallis entropy for negative values of $q$ does not fulfill Kolmogorov axioms defined by Abe \cite{abe97}. This can be overcome by assuming only $q \in [0,2]$.

It is also possible to obtain the multiplicative duality, when we allow $q$-dependent transformation $\alpha(q)$. In this case, we simply choose $\alpha(q) = -q$, which results into
\begin{equation}
q_{-q} = \frac{q-q-1}{-q} = 1/q\, .
\end{equation}
Both additive and multiplicative dualities have been recently discussed e.g. in Ref. \cite{tsallis17}.

\section{Applications of Tsallis parameter transformation in thermostatistics}\label{sec:ap}
In order to understand the physical interpretation of the transformation $S_q \rightarrow S_{q_\alpha}$, let us focus on the case of finite heat bath. For this system is
\begin{equation}
q(N)_\alpha = \frac{q(N)+ \alpha-1}{\alpha} = \frac{N + (\alpha-1)(N-1)}{\alpha(N-1)} = \frac{\alpha(N-1)+1}{\alpha(N-1)}\, .
\end{equation}
Therefore, parameter $q_\alpha$ corresponds to a system in contact with a finite heat bath consisting of $N_\alpha = \alpha(N-1)+1$ particles, from which we get that
 \begin{equation}
(N_\alpha -1) = \alpha (N-1).
\end{equation}
Thus, transformation $q \rightarrow q_\alpha$ corresponds to rescaling the number of particles in the bath. Generally, the transform describes the shift between classes of Tsallis $q$-dditivity. For the case of finite heat bath, we always have $q(N)_\alpha > 1$ for $\alpha > 0$.

Rescaling the number of particles in the finite heat bath is one of the physical applications of the nonadditivity parameter transformation. On the other hand, it is possible to find a nice physical interpretation of the transformation for systems with temperature fluctuations. Such systems have been investigated by Beck in Ref. \cite{beck02} followed by several other authors.  In a general system in contact with a heat bath with temperature fluctuations it is possible to express the non-additivity parameter $q$ as \cite{biro14}
\begin{equation}
q = 1 - \frac{1}{C} + \frac{\Delta \beta^2}{\langle \beta \rangle^2}
\end{equation}
where $C$ is the heat capacity of the reservoir and $\frac{\Delta \beta^2}{\langle \beta \rangle^2}$ is the relative temperature fluctuation. Let us note for the case of finite heat reservoir discussed in the previous section, the heat capacity is negative, as discussed e.g. in Ref. \cite{bagci13}. On the other hand, for positive heat capacity, it is possible to reach the region $q<1$ and $q=1$ determines $\frac{1}{\sqrt{C}} = \frac{\Delta \beta}{\langle \beta \rangle}$. For systems, with large fluctuations, i.e. $\frac{\Delta \beta^2}{\langle \beta \rangle^2} \gg \frac{1}{C}$, we can neglect $1/C$. Then, for the system with rescaled nonadditivity parameter $q_\alpha$, we have
\begin{equation}
\frac{\Delta \beta_\alpha^2}{\langle \beta_\alpha \rangle^2} = q_\alpha- 1 = \frac{q-1}{\alpha} = \frac{1}{\alpha} \frac{\Delta \beta^2}{\langle \beta \rangle^2}\, .
\end{equation}
Thus, rescaling the nonadditivity parameter also rescales the relative fluctuations in the system.

Finally, a nice application of the nonadditivity parameter transformation is the \emph{quasi-additivity} rule for Tsallis entropy for $q$ close to one \cite{beck02}. In this case, it is possible to make the expansion of $\sum_i= p_i^q$ as
\begin{equation}
\sum_i p_i^q =  \sum_i p_i e^{(q-1)\log p_i} = 1 + (q-1) \sum_i p_i \log p_i + \frac{(q-1)^2}{2} \sum_i p_i (\log p_i)^2 + \dots
\end{equation}
In this approximation it is possible to find a quasi-additivity rule for Tsallis entropy, which can be expressed as
\begin{equation}\label{eq:tsallis}
S_q(P_A) + S_q(P_B) = S_{q_\alpha}(P_{A \cup B})
\end{equation}
with appropriate $\alpha$. For $A=B$, the left-hand side is equal to
\begin{equation}
2 S_q(P_A) = \frac{2}{q-1} \left(1-\sum_i p_i^q\right) = - 2 \sum_i p_i \log p_i - (q-1) \sum_i p_i (\log p_i)^2 + \dots
\end{equation}
while the right-hand side can be expressed as
\begin{eqnarray}
S_{q_\alpha}(P_{A^2}) = \frac{1}{q_\alpha-1}(1-\sum_{ij} p_{ij}^{q_\alpha}) = \frac{1}{q_\alpha-1}(1-(\sum_{i} p_{i}^{q_\alpha})^2)\nonumber\\
 = -2 \sum_i p_i \log p_i  - (q_\alpha-1)\left[(\sum_i p_i \log p_i)^2 + \sum_i p_i (\log p_i)^2\right] + \dots
\end{eqnarray}
From this, we can determine $\alpha$ as
\begin{equation}
\alpha = \frac{q-1}{q_\alpha-1} = 1 + \frac{\langle I(P) \rangle^2}{\langle I(P)^2 \rangle} 
\end{equation}
where $I(P) = \log \frac{1}{p}$ is the Hartley information. $\langle I(P) \rangle$ is nothing else than Shannon entropy and $\langle I(P)^2 \rangle$ represents the fluctuation of the information. It is clear from Jensen's inequality that $\alpha \in [1,2]$, where $\alpha=1$ if $p_i = \delta_{ij}$, i.e. distribution of a certain event, and $\alpha=2$ for $p_i = 1/n$, i.e. uniform distribution. 
%Generally, the $q_\alpha$ transform describes rescaling of the system.
\section{Generalized distributive law in $q$-deformed algebra}\label{sec:qa}
Borges \cite{borges04} introduced a deformed calculus inspired by the nonadditivity rule of Tsallis entropy. This extends ordinary operators and certain functions to the nonadditive, $q$-deformed regime. Let us remind the definitions $q$-deformed operators:
\begin{align}
x \oplus_q y = x + y + (1-q) x y\\
x \ominus_q y = \frac{x-y}{1+(1-q)y}\\
x \otimes_q y = [x^{1-q}+y^{1-q}-1]^{1/(1-q)}\\
x \oslash_q y = [x^{1-q}-y^{1-q}+1]^{1/(1-q)}
\end{align}
which for $q \rightarrow 1$ become ordinary operators, i.e. ordinary addition, subtraction, multiplication and division. Naturally, Tsallis entropy of a compound system can be expressed as the $q$-sum of entropies corresponding to the subsystems (Eq.~\ref{eq:tsallis}). The operators are commutative and associative, but they are not distributive. On the other hand, it is possible to recover a generalized distributivity which relates $q$-operation with $q_\alpha$-operation. For example, we can express the $q$-addition distributivity law as
\begin{equation*}
\alpha (x \oplus_{q} y) = \alpha x + \alpha y + \alpha (1-q) xy = \alpha x + \alpha y + \frac{1-q}{\alpha} (\alpha x)(\alpha y) = (\alpha x) \oplus_{q_\alpha} (\alpha y).
\end{equation*}
Therefore, we obtain that
\begin{align}
\alpha (x \oplus_q y) = (\alpha x) \oplus_{q_\alpha} (\alpha y)\\
\alpha(x \ominus_q y) = (\alpha x) \ominus_{q_\alpha} (\alpha y)\\
(x \otimes_q y)^\alpha = (x^\alpha) \otimes_{q_\alpha} (y^\alpha)\\
(x \oslash_q y)^\alpha =  (x^\alpha) \oslash_{q_\alpha} (y^\alpha).
\end{align}
It is also possible to define $q$-deformed versions of exponential and logarithmic functions:
\begin{align}
\exp_q x = [1+(1-q)x]^{1/(1-q)}\\
\ln_q x = \frac{x^{1-q}-1}{1-q}.
\end{align}
Naturally $\ln_q (\exp_q x) = \exp_q (\ln_q x) = x$ for appropriate $x$. Moreover, they are tightly related to $q$-deformed addition and multiplication. The relations can be summarized by following relations
\begin{align}
e_q(x) e_q(y) = e_q(x \oplus_q y)\\
e_q(x+y) = e_q(x) \otimes_q e_q(y)\\
\ln_q(xy)= \ln_q(x) \oplus_q \ln_q(y)\\
\ln_q(x)+\ln_q(y) = \ln_q(x \otimes_q y)\, .
\end{align}
From the properties of $q_\alpha$ operators, is possible to derive further relations generalizing ordinary properties of exponential and logarithmic functions
\begin{align}
(\exp_{q}x)^{\alpha} = \exp_{q_\alpha}(\alpha x)\\
\alpha \ln_{q}(x) = \ln_{q_\alpha} (x^{\alpha}).
\end{align}
Again, for $q \rightarrow 1$ become all functions ordinary logartihms and exponentials. Generally, the transformation $q_\alpha$ appears naturally in $q$-deformed calculus in connection with rescaling.

\section{MaxEnt distributions of Tsallis entropy with rescaled nonadditivity parameter}\label{sec:me}
The important application of entropy in statistical physics is the Maximum entropy principle (MaxEnt principle) proposed by Jaynes \cite{jaynes57}. It states that the realized distribution (called MaxEnt distribution) of a statistical system maximizes the information uncertainty (quantified by the entropy functional) under given constraints. In connection with Tsallis entropy, the usual distribution obtained by MaxEnt procedure is the $q$-gaussian distribution
\begin{equation}
p_q(E) \propto \exp_q(-\Omega \Delta E).
\end{equation}
Tsallis entropy is often used as a tool for obtaining $q$-gaussian distribution. On the other hand, it is necessary to say that particular form of MaxEnt distribution always depends on the entropy functional as well as on exact form of constraints. It has been shown that we can obtain many different types of MaxEnt distributions from the same entropy functional under different constraints (for Shannon entropy, see e.g. \cite{lisman72,park09}). For example, Bercher showed that the $q$-gaussian distribution can be obtained from Shannon entropy under special ``tail'' constraints \cite{bercher08}.

In the framework of Tsallis statistics, the energy constraint is often formulated in the form of escort mean
\begin{equation}
\langle E \rangle_r = \sum_k \rho_k(r) E_k
\end{equation}
where $\rho_k(r) = p_k^r /\sum_j p_j^r$ is the escort distribution, introduced in connection with dynamic systems \cite{beck93} and further investigated e.g. in \cite{beck2004}. Its application to thermodynamics was pointed out in connection with superstatistics \cite{tsallis03}. For Tsallis entropy $S_q$, there are two typical choices of constraint parameter $r$, i.e. linear constraints $r=1$, and $q$-escort constraints, i.e. $r=q$. Both choices lead to $q$-gaussian distribution. The particular form of constraints was extensively discussed by several authors~\cite{tsallis98,rama00,abe01,plastino01}. We choose the \mbox{$q$-escort} constraints, so $r=q$. For the case when $r=1$, the rescaling does not affect the functional form of MaxEnt distributions, so we get $q_\alpha$-gaussian distributions.

Let us investigate the MaxEnt distribution of Tsallis entropy with rescaled nonadditivity parameter $S_{q_\alpha}$ under constraints $\sum_k p_k =1$ and $\langle E \rangle_q =\mathcal{E}_q$. This means to extremize the functional
\begin{equation}
L_{q,\alpha}(P) = S_{q_\alpha}(P) - \Phi \sum_k p_k - \Omega \sum_k \rho_k(q) E_k
\end{equation}
which leads to the equation $\partial L_{q,\alpha}(P)/\partial p_i = 0$ which can be expressed as
\begin{equation}\label{eq:maxent}
\frac{q_\alpha}{1-q_\alpha} \, p_i^{(q-1)/\alpha} - \Phi -   \frac{q \Omega \Delta E}{Z_q} \, p_i^{q-1}   = 0\, ,
\end{equation}
where $\Delta E =(E_i - \mathcal{E}_q)$ and $Z_q = \sum_k p_k^q$. Lagrange parameter $\Phi$ can be determined by multiplying the equation by $p_i$ and summing over $i$. We obtain that $\Phi = \frac{q_\alpha}{1-q_\alpha} Z_{q_\alpha}$. Eq. \eqref{eq:maxent} is a trinomial equation, possibly of  non-natural order. Naturally, it boils down to linear equation for $\alpha=1$. After a substitution, we end with equation
\begin{equation}\label{eq:tri}
1 - x + b x^{\alpha} = 0\, ,
\end{equation}
where $x = p_i^{(q-1)/\alpha}/Z_{q_\alpha}$ and $b= \frac{q(1-q)}{q+\alpha-1} \frac{Z_{q_\alpha}^{\alpha-1}}{Z_q} \,  \, \Omega \Delta E$. The root which for $b \rightarrow 0$ reduces to 1 can be found in terms of infinite series as \cite{glasser00}
\begin{equation}\label{eq:sum}
x = 1 + \sum_{n=1}^{\infty} \binom{\alpha n}{n-1} \frac{b^n}{n}\, .
\end{equation}
Thus, Eq.~\ref{eq:tri} has a positive real solution at least for sufficiently small $\Delta E$. In several cases, it is possible to express the root analytically:
\begin{itemize}
  \item $\alpha=1$, $q_1 = 1$: the series in Eq.~\ref{eq:sum} becomes ordinary geometric series, so $x= 1/(1-b)$. This corresponds to the fact that Eq.~\ref{eq:tri} becomes linear equation. The resulting distribution is therefore ordinary $q$-gaussian distribution.

  \item $\alpha=1/2$, $q_{1/2} = 2q-1$: we get a quadratic equation in terms of $\sqrt{x}$. The root can be expressed as $x=1+ \frac{1}{2} b_{1/2} \, (b_{1/2}+ \sqrt{4+b_{1/2}^2})$, where $b_{1/2} =  \frac{2q(1-q)}{2q-1} \frac{1}{Z_{q_{1/2}}^{1/2} Z_q}\, \Omega \Delta E$.
  The resulting probability distribution can be expressed as
  \begin{equation}p_{q,1/2}(E) \propto \left[1+\frac{C_q^2}{2}\Delta E^2 \left(1+ \sqrt{1+4/(C_q \Delta E^2)}\right)\right]^{\frac{1}{2(q-1)}}\end{equation}
  where $C_q$ contains all energy-independent terms.

  \item $\alpha=2$, $q_2 = (q+1)/2$: Eq. \ref{eq:tri} becomes quadratic equation, whose roots are $x=\frac{1 \pm \sqrt{1-4 b_2}}{2 b_2}$, where $b_2 = \frac{q(1-q)}{1+q} \frac{Z_{q_2}}{Z_q} \, \Omega \Delta E$. Thus, it is possible to express the resulting distribution as
      \begin{equation}
      p_{q,2}(E) \propto \left(\frac{1 \pm \sqrt{1-4 c_q \Delta E}}{2 c_Q \Delta E}\right)^{\frac{2}{q-1}}
      \end{equation}
      where $c_q$ contains all energy-independent terms. Interestingly, we can express the relation between $Z_{q_2}$ and $Z_q$ by the Cauchy-Schwarz inequality as
  \begin{equation}
  Z_{q_2} = Z_{(q+1)/2} \leq \sqrt{Z_q} \sqrt{Z_1} = \sqrt{Z_q}.
  \end{equation}

  \item $\alpha \rightarrow \infty$, $q_{\infty} = 1$: in the limit case we recover the ordinary additivity resulting into Shannon entropy $S_1(P)=-\sum_i p_i \ln p_i$. Therefore, Eq.~\ref{eq:maxent} becomes
  \begin{equation}
   - \ln p_i + S_1(P) - \frac{q \Omega \Delta E}{Z_q} \, p_i^{q-1} = 0.
  \end{equation}
 It is possible to find the solution in terms of Lambert W-function \cite{corless96} as
 \begin{equation}p_{q,\infty}(E) = \exp\left[S_1 + \frac{1}{q-1}\,  W\left( \frac{ (q-1) q e^{(q-1)S_1} \Omega \Delta E }{Z_q} \right) \right]. \end{equation}
 Lambert W-function has found many applications in MaxEnt distributions \cite{jizba16,hanel11} as well as in other physical applications \cite{valluri00}.
\end{itemize}
Additionally, for $\alpha=n$, where $n \in \mathds{N}$, we obtain the trinomial equation of order $n$, whose root can be expressed as a finite sum of at most $n-2$ hypergeometric functions \cite{glasser00,szabo10}. Generally, for each $\alpha$ we get a special class of ``$q$-deformed exponentials'', which for $q \rightarrow 1$ boils down to ordinary exponential distributions.

Interestingly, we get almost the same results when we replace Tsallis entropy by additive R\'{e}nyi entropy. R\'{e}nyi entropy can be used in description of multifractal systems \cite{jizba04,jizba14} as well in statistical physics \cite{jizba04a,lenzi2000}. It is defined a $R_q(P) = \frac{\ln Z_q}{1-q}$, so it is possible to find the relation between both entropies, which is $R_q = \ln \exp_q S_q$. The corresponding MaxEnt distributions differ only in the presence of $Z_q$ in the first term of Eq. \ref{eq:maxent}, because
\begin{equation}\frac{\partial R_q}{\partial p_i} = \frac{q p_i^{q-1}}{(1-q) Z_q}.\end{equation}

\section{Note on Jizba-Arimitsu Hybrid entropy}\label{sec:ja}
Jizba-Arimitsu Hybrid entropy was defined as the overlap between Tsallis entropy and R\'{e}nyi entropy \cite{arimitsu04} and its properties have been recently discussed in Refs.~\cite{jizba16,ilic17,jizba17a,cankaya17}. Similarly to Tsallis entropy, it follows $q$-deformed additivity rule for independent events. On the other hand, corresponding conditional entropy is defined in the way of R\'{e}nyi entropy, i.e. in terms of generalized Kolmogorov-Nagumo mean. It can be expressed as
\begin{equation}
D_q(P) = \ln_q \exp\left( - \sum_i \rho_i(q) \ln p_i\right)\, .
\end{equation}
In Ref.~\cite{jizba16} was shown that the hybrid entropy is properly defined only for $q \geq 1/2$, while for $q < 1/2$ it does not satisfy the maximality axiom. On the other hand, Tsallis entropy and $q$-deformed calculus are defined for all positive values. In Ref.~\cite{cankaya17} was shown by methods of information geometry that it is possible to fix this issue by introducing the Average Hybrid entropy $A_q$, which can be defined as $A_q(P) = D_{\frac{q+1}{2}}(P)$. We show that it is possible to arrive at this result from the view of nonadditivity parameter transformation. Let us assume that the hybrid entropy is the entropy with rescaled nonadditivity index $q_\alpha$. We find such $\alpha$, for which
\begin{equation}
0_\alpha = \frac{\alpha-1}{\alpha} =  \frac{1}{2}\,
\end{equation}
which results into $\alpha=2$, corresponding to $q_2 =\frac{q+1}{2}$. This result is in agreement with results from \cite{cankaya17}.

\section{Conclusions}
Nonadditivity parameter transformation group represents the way of switching between particular nonadditive regimes in systems with different types of Tsallis nonadditivity, which characterized by the quadratic term $(1-q)xy$ added to ordinary additivity $x+y$. This transformation corresponds to a situation, when the source of nonadditivity is rescaled, e.g. the number of particles in the finite heat bath or relative temperature fluctuations. Typically, Tsallis entropy describes systems which state space grows polynomially and the transformation describes the shift of the scaling exponent. The main aim of the paper was to introduce the transformation group and discuss its main properties and possible applications. From the mathematical point of view, the group is closely related to $q$-deformed algebra and provides a way how to generalize distributive law known from ordinary algebra. We have introduced a class of MaxEnt distributions which were obtained from Tsallis entropies with rescaled nonadditivity parameter under escort energy constraints. For each scaling factor $\alpha$, resulting distributions provide a class of $q$-deformed distributions. Moreover, we have briefly discussed the relation of the nonadditivity parameter transformation group to average hybrid entropy, which is a version of Jizba-Arimitsu Hybrid entropy with rescaled nonadditivity parameter. Naturally, the group transformation is closely related to $q$-deformed additivity and Tsallis entropy. The existence of such group for other types of nonadditivity and its properties are the subject of a future research.

\section*{Acknowledgements}
I would like to thank the three anonymous reviewers for their helpful and constructive comments. I acknowledge the financial support from the Czech Science Foundation, grant No. 17-33812L.
\section*{References}
\nolinenumbers
\bibliography{references}

\begin{thebibliography}{53}
\expandafter\ifx\csname natexlab\endcsname\relax\def\natexlab#1{#1}\fi
\providecommand{\url}[1]{\texttt{#1}}
\providecommand{\href}[2]{#2}
\providecommand{\path}[1]{#1}
\providecommand{\DOIprefix}{doi:}
\providecommand{\ArXivprefix}{arXiv:}
\providecommand{\URLprefix}{URL: }
\providecommand{\Pubmedprefix}{pmid:}
\providecommand{\doi}[1]{\href{http://dx.doi.org/#1}{\path{#1}}}
\providecommand{\Pubmed}[1]{\href{pmid:#1}{\path{#1}}}
\providecommand{\bibinfo}[2]{#2}
\ifx\xfnm\relax \def\xfnm[#1]{\unskip,\space#1}\fi
%Type = Article
\bibitem[{Shannon(1948)}]{shannon48}
\bibinfo{author}{C.~Shannon}, \bibinfo{journal}{Bell System Technical Journal}
  \bibinfo{volume}{27} (\bibinfo{year}{1948}) \bibinfo{pages}{379--423}.
%Type = Book
\bibitem[{R{\'e}nyi(1976)}]{renyi76}
\bibinfo{author}{A.~R{\'e}nyi}, \bibinfo{title}{Selected Papers of Alfr{\'e}d
  R{\'e}nyi}, number~\bibinfo{number}{2} in \bibinfo{series}{Selected Papers of
  Alfr{\'e}d R{\'e}nyi}, \bibinfo{publisher}{Akad{\'e}miai Kiad{\'o}},
  \bibinfo{year}{1976}.
%Type = Article
\bibitem[{Tsallis(1988)}]{tsallis88}
\bibinfo{author}{C.~Tsallis}, \bibinfo{journal}{Journal of Statistical Physics}
  \bibinfo{volume}{52} (\bibinfo{year}{1988}) \bibinfo{pages}{479--487}.
%Type = Article
\bibitem[{Abe(1997)}]{abe97}
\bibinfo{author}{S.~Abe}, \bibinfo{journal}{Phys. Lett. A}
  \bibinfo{volume}{224} (\bibinfo{year}{1997}) \bibinfo{pages}{326--330}.
%Type = Article
\bibitem[{Havrda and Charv\'{a}t(1967)}]{havrda67}
\bibinfo{author}{J.~Havrda}, \bibinfo{author}{F.~Charv\'{a}t},
  \bibinfo{journal}{Kybernetika} \bibinfo{volume}{3} (\bibinfo{year}{1967})
  \bibinfo{pages}{30--35}.
%Type = Article
\bibitem[{Frank and Daffertshofer(2000)}]{frank20}
\bibinfo{author}{T.~Frank}, \bibinfo{author}{A.~Daffertshofer},
  \bibinfo{journal}{Physica A} \bibinfo{volume}{285} (\bibinfo{year}{2000})
  \bibinfo{pages}{351--366}.
%Type = Article
\bibitem[{Sharma et~al.(1978)Sharma, Mitter, and Mohan}]{sharma78}
\bibinfo{author}{B.~D. Sharma}, \bibinfo{author}{J.~Mitter},
  \bibinfo{author}{M.~Mohan}, \bibinfo{journal}{Information and Control}
  \bibinfo{volume}{39} (\bibinfo{year}{1978}) \bibinfo{pages}{323--336}.
%Type = Article
\bibitem[{Bercher(2012)}]{bercher12}
\bibinfo{author}{J.-F. Bercher}, \bibinfo{journal}{Physica A}
  \bibinfo{volume}{391} (\bibinfo{year}{2012}) \bibinfo{pages}{4460--4469}.
%Type = Article
\bibitem[{Jizba and Arimitsu(2004)}]{arimitsu04}
\bibinfo{author}{P.~Jizba}, \bibinfo{author}{T.~Arimitsu},
  \bibinfo{journal}{Physica A} \bibinfo{volume}{340} (\bibinfo{year}{2004})
  \bibinfo{pages}{110--116}.
%Type = Article
\bibitem[{Hanel and Thurner(2011)}]{hanel11}
\bibinfo{author}{R.~Hanel}, \bibinfo{author}{S.~Thurner},
  \bibinfo{journal}{Europhys. Lett.} \bibinfo{volume}{93}
  (\bibinfo{year}{2011}) \bibinfo{pages}{20006}.
%Type = Article
\bibitem[{Tempesta(2011)}]{tempesta11}
\bibinfo{author}{P.~Tempesta}, \bibinfo{journal}{Phys. Rev. E}
  \bibinfo{volume}{84} (\bibinfo{year}{2011}) \bibinfo{pages}{021121}.
%Type = Article
\bibitem[{Ili{\'c} and Stankovi{\'c}(2014)}]{ilic14}
\bibinfo{author}{V.~M. Ili{\'c}}, \bibinfo{author}{M.~S. Stankovi{\'c}},
  \bibinfo{journal}{Physica A} \bibinfo{volume}{411} (\bibinfo{year}{2014})
  \bibinfo{pages}{138--145}.
%Type = Article
\bibitem[{Bir{\'o} et~al.(2015)Bir{\'o}, Barnaf{\"o}ldi, and V{\'a}n}]{biro15}
\bibinfo{author}{T.~Bir{\'o}}, \bibinfo{author}{G.~Barnaf{\"o}ldi},
  \bibinfo{author}{P.~V{\'a}n}, \bibinfo{journal}{Physica A}
  \bibinfo{volume}{417} (\bibinfo{year}{2015}) \bibinfo{pages}{215--220}.
%Type = Article
\bibitem[{Jizba and Korbel(2016)}]{jizba16}
\bibinfo{author}{P.~Jizba}, \bibinfo{author}{J.~Korbel},
  \bibinfo{journal}{Physica A} \bibinfo{volume}{444} (\bibinfo{year}{2016})
  \bibinfo{pages}{808--827}.
%Type = Article
\bibitem[{Plastino and Plastino(1994)}]{plastino94}
\bibinfo{author}{A.~Plastino}, \bibinfo{author}{A.~Plastino},
  \bibinfo{journal}{Phys. Lett. A} \bibinfo{volume}{193} (\bibinfo{year}{1994})
  \bibinfo{pages}{140 -- 143}.
%Type = Article
\bibitem[{Adib et~al.(2003)Adib, Moreira, Andrade~Jr, and Almeida}]{adib03}
\bibinfo{author}{A.~B. Adib}, \bibinfo{author}{A.~A. Moreira},
  \bibinfo{author}{J.~S. Andrade~Jr}, \bibinfo{author}{M.~P. Almeida},
  \bibinfo{journal}{Physica A} \bibinfo{volume}{322} (\bibinfo{year}{2003})
  \bibinfo{pages}{276--284}.
%Type = Article
\bibitem[{Kaniadakis(2001)}]{kaniadakis01}
\bibinfo{author}{G.~Kaniadakis}, \bibinfo{journal}{Physica A}
  \bibinfo{volume}{296} (\bibinfo{year}{2001}) \bibinfo{pages}{405--425}.
%Type = Article
\bibitem[{Plastino et~al.(2014)Plastino, Curado, and Nobre}]{plastino14}
\bibinfo{author}{A.~Plastino}, \bibinfo{author}{E.~Curado},
  \bibinfo{author}{F.~Nobre}, \bibinfo{journal}{Physica A}
  \bibinfo{volume}{403} (\bibinfo{year}{2014}) \bibinfo{pages}{13 -- 20}.
%Type = Article
\bibitem[{Jizba et~al.(2017)Jizba, Korbel, and Zatloukal}]{jizba17}
\bibinfo{author}{P.~Jizba}, \bibinfo{author}{J.~Korbel},
  \bibinfo{author}{V.~Zatloukal}, \bibinfo{journal}{Phys. Rev. E}
  \bibinfo{volume}{95} (\bibinfo{year}{2017}) \bibinfo{pages}{022103}.
%Type = Article
\bibitem[{Borges(2004)}]{borges04}
\bibinfo{author}{E.~P. Borges}, \bibinfo{journal}{Physica A}
  \bibinfo{volume}{340} (\bibinfo{year}{2004}) \bibinfo{pages}{95--101}.
%Type = Article
\bibitem[{Nivanen et~al.(2003)Nivanen, Le~Mehaute, and Wang}]{nivanen03}
\bibinfo{author}{L.~Nivanen}, \bibinfo{author}{A.~Le~Mehaute},
  \bibinfo{author}{Q.~Wang}, \bibinfo{journal}{Reports on Mathematical Physics}
  \bibinfo{volume}{52} (\bibinfo{year}{2003}) \bibinfo{pages}{437--444}.
%Type = Article
\bibitem[{Kalogeropoulos(2005)}]{kalogeropoulos05}
\bibinfo{author}{N.~Kalogeropoulos}, \bibinfo{journal}{Physica A}
  \bibinfo{volume}{356} (\bibinfo{year}{2005}) \bibinfo{pages}{408--418}.
%Type = Article
\bibitem[{Bercher(2012)}]{bercher2012a}
\bibinfo{author}{J.-F. Bercher}, \bibinfo{journal}{J. Phys. A}
  \bibinfo{volume}{45} (\bibinfo{year}{2012}) \bibinfo{pages}{255303}.
%Type = Article
\bibitem[{Bercher(2013)}]{bercher13}
\bibinfo{author}{J.-F. Bercher}, \bibinfo{journal}{Physica A}
  \bibinfo{volume}{392} (\bibinfo{year}{2013}) \bibinfo{pages}{3140--3154}.
%Type = Article
\bibitem[{{\c{C}}ankaya and Korbel(2017)}]{cankaya17}
\bibinfo{author}{M.~N. {\c{C}}ankaya}, \bibinfo{author}{J.~Korbel},
  \bibinfo{journal}{Physica A} \bibinfo{volume}{475} (\bibinfo{year}{2017})
  \bibinfo{pages}{1--10}.
%Type = Article
\bibitem[{Sicuro et~al.(2016)Sicuro, Bagchi, and Tsallis}]{sicuro2016}
\bibinfo{author}{G.~Sicuro}, \bibinfo{author}{D.~Bagchi},
  \bibinfo{author}{C.~Tsallis}, \bibinfo{journal}{Phys. Lett. A}
  \bibinfo{volume}{380} (\bibinfo{year}{2016}) \bibinfo{pages}{2025 -- 2030}.
%Type = Article
\bibitem[{Bagchi and Tsallis(2017)}]{bagchi17}
\bibinfo{author}{D.~Bagchi}, \bibinfo{author}{C.~Tsallis},
  \bibinfo{journal}{Phys. Lett. A} \bibinfo{volume}{381} (\bibinfo{year}{2017})
  \bibinfo{pages}{1123--1128}.
%Type = Article
\bibitem[{Hanel and Thurner(2011)}]{thurner11}
\bibinfo{author}{R.~Hanel}, \bibinfo{author}{S.~Thurner},
  \bibinfo{journal}{Europhys. Lett.} \bibinfo{volume}{96}
  (\bibinfo{year}{2011}) \bibinfo{pages}{50003}.
%Type = Article
\bibitem[{Tsallis(2017)}]{tsallis17}
\bibinfo{author}{C.~Tsallis}, \bibinfo{journal}{The European Physical Journal
  Special Topics} \bibinfo{volume}{226} (\bibinfo{year}{2017})
  \bibinfo{pages}{455--466}.
%Type = Article
\bibitem[{Beck(2002)}]{beck02}
\bibinfo{author}{C.~Beck}, \bibinfo{journal}{EPL (Europhysics Letters)}
  \bibinfo{volume}{57} (\bibinfo{year}{2002}) \bibinfo{pages}{329}.
%Type = Article
\bibitem[{Biró et~al.(2014)Biró, Ván, Barnaföldi, and Ürmössy}]{biro14}
\bibinfo{author}{T.~S. Biró}, \bibinfo{author}{P.~Ván},
  \bibinfo{author}{G.~G. Barnaföldi}, \bibinfo{author}{K.~Ürmössy},
  \bibinfo{journal}{Entropy} \bibinfo{volume}{16} (\bibinfo{year}{2014})
  \bibinfo{pages}{6497--6514}. \URLprefix
  \url{http://www.mdpi.com/1099-4300/16/12/6497}.
%Type = Article
\bibitem[{Bagci and Oikonomou(2013)}]{bagci13}
\bibinfo{author}{G.~B. Bagci}, \bibinfo{author}{T.~Oikonomou},
  \bibinfo{journal}{Phys. Rev. E} \bibinfo{volume}{88} (\bibinfo{year}{2013})
  \bibinfo{pages}{042126}.
%Type = Article
\bibitem[{Jaynes(1957)}]{jaynes57}
\bibinfo{author}{E.~T. Jaynes}, \bibinfo{journal}{Phys. Rev.}
  \bibinfo{volume}{106} (\bibinfo{year}{1957}) \bibinfo{pages}{620--630}.
%Type = Article
\bibitem[{Lisman and Van~Zuylen(1972)}]{lisman72}
\bibinfo{author}{J.~Lisman}, \bibinfo{author}{M.~Van~Zuylen},
  \bibinfo{journal}{Statistica Neerlandica} \bibinfo{volume}{26}
  (\bibinfo{year}{1972}) \bibinfo{pages}{19--23}.
%Type = Article
\bibitem[{Park and Bera(2009)}]{park09}
\bibinfo{author}{S.~Y. Park}, \bibinfo{author}{A.~K. Bera},
  \bibinfo{journal}{Journal of Econometrics} \bibinfo{volume}{150}
  (\bibinfo{year}{2009}) \bibinfo{pages}{219--230}.
%Type = Article
\bibitem[{Bercher(2008)}]{bercher08}
\bibinfo{author}{J.-F. Bercher}, \bibinfo{journal}{Phys. Lett. A}
  \bibinfo{volume}{372} (\bibinfo{year}{2008}) \bibinfo{pages}{5657--5659}.
%Type = Book
\bibitem[{Beck and Schl\"ogl(1993)}]{beck93}
\bibinfo{author}{C.~Beck}, \bibinfo{author}{F.~Schl\"ogl},
  \bibinfo{title}{Thermodynamics of chaotic systems. An introduction.},
  \bibinfo{publisher}{Cambridge Universtiy Press}, \bibinfo{year}{1993}.
%Type = Article
\bibitem[{Beck(2004)}]{beck2004}
\bibinfo{author}{C.~Beck}, \bibinfo{journal}{Physica A} \bibinfo{volume}{342}
  (\bibinfo{year}{2004}) \bibinfo{pages}{139--144}.
%Type = Article
\bibitem[{Tsallis and Souza(2003)}]{tsallis03}
\bibinfo{author}{C.~Tsallis}, \bibinfo{author}{A.~M. Souza},
  \bibinfo{journal}{Phys. Rev. E} \bibinfo{volume}{67} (\bibinfo{year}{2003})
  \bibinfo{pages}{026106}.
%Type = Article
\bibitem[{Tsallis et~al.(1998)Tsallis, Mendes, and Plastino}]{tsallis98}
\bibinfo{author}{C.~Tsallis}, \bibinfo{author}{R.~Mendes},
  \bibinfo{author}{A.~Plastino}, \bibinfo{journal}{Physica A: Statistical
  Mechanics and its Applications} \bibinfo{volume}{261} (\bibinfo{year}{1998})
  \bibinfo{pages}{534 -- 554}.
%Type = Article
\bibitem[{Rama(2000)}]{rama00}
\bibinfo{author}{S.~K. Rama}, \bibinfo{journal}{Phys. Lett. A}
  \bibinfo{volume}{276} (\bibinfo{year}{2000}) \bibinfo{pages}{103--108}.
%Type = Article
\bibitem[{Abe et~al.(2001{\natexlab{a}})Abe, Mart{\i}nez, Pennini, and
  Plastino}]{abe01}
\bibinfo{author}{S.~Abe}, \bibinfo{author}{S.~Mart{\i}nez},
  \bibinfo{author}{F.~Pennini}, \bibinfo{author}{A.~Plastino},
  \bibinfo{journal}{Phys. Lett. A} \bibinfo{volume}{281}
  (\bibinfo{year}{2001}{\natexlab{a}}) \bibinfo{pages}{126--130}.
%Type = Article
\bibitem[{Abe et~al.(2001{\natexlab{b}})Abe, Mart{\i}nez, Pennini, and
  Plastino}]{plastino01}
\bibinfo{author}{S.~Abe}, \bibinfo{author}{S.~Mart{\i}nez},
  \bibinfo{author}{F.~Pennini}, \bibinfo{author}{A.~Plastino},
  \bibinfo{journal}{Phys. Lett. A} \bibinfo{volume}{278}
  (\bibinfo{year}{2001}{\natexlab{b}}) \bibinfo{pages}{249--254}.
%Type = Article
\bibitem[{Glasser(2000)}]{glasser00}
\bibinfo{author}{M.~Glasser}, \bibinfo{journal}{Journal of Computational and
  Applied Mathematics} \bibinfo{volume}{118} (\bibinfo{year}{2000})
  \bibinfo{pages}{169--173}.
%Type = Article
\bibitem[{Corless et~al.(1996)Corless, Gonnet, Hare, Jeffrey, and
  Knuth}]{corless96}
\bibinfo{author}{R.~M. Corless}, \bibinfo{author}{G.~H. Gonnet},
  \bibinfo{author}{D.~E. Hare}, \bibinfo{author}{D.~J. Jeffrey},
  \bibinfo{author}{D.~E. Knuth}, \bibinfo{journal}{Advances in Computational
  mathematics} \bibinfo{volume}{5} (\bibinfo{year}{1996})
  \bibinfo{pages}{329--359}.
%Type = Article
\bibitem[{Valluri et~al.(2000)Valluri, Jeffrey, and Corless}]{valluri00}
\bibinfo{author}{S.~R. Valluri}, \bibinfo{author}{D.~J. Jeffrey},
  \bibinfo{author}{R.~M. Corless}, \bibinfo{journal}{Canadian Journal of
  Physics} \bibinfo{volume}{78} (\bibinfo{year}{2000})
  \bibinfo{pages}{823--831}.
%Type = Article
\bibitem[{Szab{\'o}(2010)}]{szabo10}
\bibinfo{author}{P.~G. Szab{\'o}}, \bibinfo{journal}{Central European Journal
  of Operations Research} \bibinfo{volume}{18} (\bibinfo{year}{2010})
  \bibinfo{pages}{97--104}.
%Type = Article
\bibitem[{Jizba and Arimitsu(2004)}]{jizba04}
\bibinfo{author}{P.~Jizba}, \bibinfo{author}{T.~Arimitsu},
  \bibinfo{journal}{Annals of Physics} \bibinfo{volume}{312}
  (\bibinfo{year}{2004}) \bibinfo{pages}{17--59}.
%Type = Article
\bibitem[{Jizba and Korbel(2014)}]{jizba14}
\bibinfo{author}{P.~Jizba}, \bibinfo{author}{J.~Korbel},
  \bibinfo{journal}{Physica A} \bibinfo{volume}{413} (\bibinfo{year}{2014})
  \bibinfo{pages}{438--458}.
%Type = Article
\bibitem[{Jizba and Arimitsu(2004)}]{jizba04a}
\bibinfo{author}{P.~Jizba}, \bibinfo{author}{T.~Arimitsu},
  \bibinfo{journal}{Phys. Rev. E} \bibinfo{volume}{69} (\bibinfo{year}{2004})
  \bibinfo{pages}{026128}.
%Type = Article
\bibitem[{Lenzi et~al.(2000)Lenzi, Mendes, and Da~Silva}]{lenzi2000}
\bibinfo{author}{E.~Lenzi}, \bibinfo{author}{R.~Mendes},
  \bibinfo{author}{L.~Da~Silva}, \bibinfo{journal}{Physica A}
  \bibinfo{volume}{280} (\bibinfo{year}{2000}) \bibinfo{pages}{337--345}.
%Type = Article
\bibitem[{Ili{\'c} and Stankovi{\'c}(2017)}]{ilic17}
\bibinfo{author}{V.~M. Ili{\'c}}, \bibinfo{author}{M.~S. Stankovi{\'c}},
  \bibinfo{journal}{Physica A} \bibinfo{volume}{466} (\bibinfo{year}{2017})
  \bibinfo{pages}{160--165}.
%Type = Article
\bibitem[{Jizba and Korbel(2017)}]{jizba17a}
\bibinfo{author}{P.~Jizba}, \bibinfo{author}{J.~Korbel},
  \bibinfo{journal}{Physica A} \bibinfo{volume}{468} (\bibinfo{year}{2017})
  \bibinfo{pages}{238 -- 243}.

\end{thebibliography}

\end{document}